# On the Support that the Special and General Theories of Relativity Provide for Rock's Argument Concerning Induced Self-Motion

Douglas M. Snyder

Los Angeles, California

## Abstract

Though Einstein and other physicists recognized the importance of an observer being at rest in an inertial reference frame for the special theory of relativity, the supporting psychological structures were not discussed much by physicists. On the other hand, Rock wrote of the factors involved in the perception of motion, including one's own motion. Rock thus came to discuss issues of significance to relativity theory, apparently without any significant understanding of how his theory might be related to relativity theory. In this paper, connections between Rock's theory on the perception of one's own motion, as well as empirical work supporting it, and relativity theory are explored.

## Text

Rock (1975/1997), following Duncker (1929/1938), discussed the concept of induced self-motion, where an individual experiences himself moving even though he is at rest. Where he experiences himself moving, and may indeed be moving as determined by an expenditure of energy (e.g., driving in a car), Rock also argued that the experienced motion may be induced. Rock maintained that the experience of motion is tied to the presence of sensory cues in an observer's environment. In the absence of these cues, or in their being discounted by an observer, the observer experiences himself at rest. For Rock, an observer's perception of his own motion includes cognitive decisions. It will be shown that the special and general theories of relativity provide support for Rock's thesis. Also, Rock's analysis of the relativity of motion provides the basis for a central concept of the special theory of relativity, namely that observers are at rest in inertial reference frames in uniform rectilinear motion relative to one another. In the general theory of relativity, Einstein essentially extended Rock's thesis on induced self-motion by discussing a situation where proprioceptive and vestibular cues that could indicate that the observer is accelerating are discounted.

# On the Support

In discussing induced self-motion, Rock (1968, 1975/1997) did not make reference to the special and general theories of relativity, and apparently he was unaware that they supported his argument concerning induced movement of the self.[1] Rock developed his argument purely on psychological grounds, based on earlier work by Duncker (1929/1938). The outline of Rock's argument will be presented, and experimental data that supports his line of reasoning will be presented. Then fundamental features of the special and general theories of relativity that support Rock's argument will be discussed.

ROCK'S ARGUMENT ON INDUCED MOTION

Rock proposed that an observer's perception that an object is moving is dependent on a change in its location that is not ascribed to the observer's movement.[2] Wallach (1959) and Rock (1975/1997) noted that there are two types of change in location that are tied to the experience of an object's motion: 1) subject-relative change and 2) object-relative change. In subject-relative change of location, the position of the observer relative to the object changes. This is often considered in terms of a change in radial direction relative to the observer. Figure 1 depicts the direction of motion of a single small circle of light against a black background. Here there is a change in the location of the circle relative to an observer, who could be the reader if the figure was the actual experimental setting.

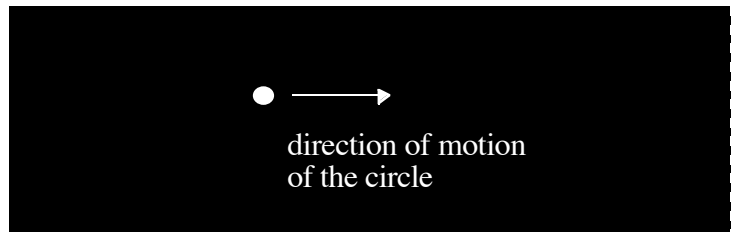

Figure 1: Motion Based on Subject-Relative Change

---

[1] In a paper on one of his experiments, Rock (1968) referred to the possibility of deducing the empirical results he obtained through a thought experiment. The term, thought experiment, was made popular by Einstein in discussing physical theory, including relativity theory. Other physicists have used the term in discussing physical theory other than that concerning relativity.

[2] Throughout this discussion of motion, it is important to realize that whether or not something is moving is relative to what is called in physics a reference frame (i.e., a spatial coordinate system, or spacetime coordinate system in the case of relativity theory) from which the state of motion of the object is being judged.

# On the Support

In object-relative change of location, the object changes its locations relative to another object, depicted in Figure 2 where two illuminated circles appear against a black background. That object-relative change of location is a distinct factor is supported by evidence that the threshold for object-relative change is smaller than that for subject-relative change. Where there is less than the minimum change in location of an object relative to an observer to effect the experience that the object is moving, an observer can still see that object in motion where this object's location changes relative to another object.

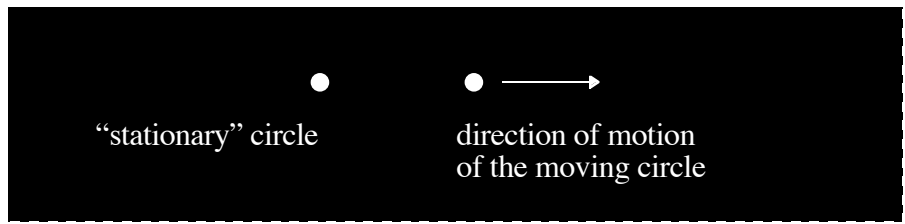

Figure 2: Motion Based on Object-Relative Change

It is possible to produce induced motion for an object where the object appears to move but does not. In this case, the motion of another object (determined for example by an expenditure of energy) relative to the object with induced motion is the basis for this apparent motion. It is possible that in certain experimental circumstances like those depicted in Figure 2, the "stationary" circle as well as the moving circle could appear to the observer to be moving in opposite directions, depicted in Figure 3. An object that is moving and that yet may appear stationary to an observer can induce motion in another object that is not moving.

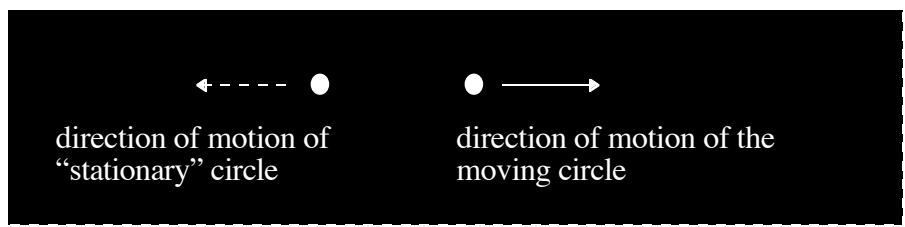

Figure 3: Induced Motion for a "Stationary" Circle

Duncker (1929/1938) and Rock (1975/1997) wrote that an object that serves as the basis for induced motion of another object is a reference frame for this latter object.[3] The former object acts as that from which the motion of an object is determined, whether or not its motion is induced or "actual."[4,5] Duncker described this relationship between a frame of reference and an object affected by it in terms of "localization" (p. 164). A reference frame may be large relative to the object in which motion is induced, and/or it may encompass it. Intensity and constancy of the object itself are also factors contributing to whether an object serves as a reference frame for another object.[6]

---

[3] The use of the term *reference frame* by Duncker and Rock is different than that found in physics.

Figure 4 depicts an example of induced motion, found in an experiment by Duncker (1929/1938), where a rectangle moves in one direction, the enclosed circle remains stationary and yet, for an observer, the circle appears to be moving in the direction opposite to that of the rectangle's motion. The rectangle remained stationary for the observer.

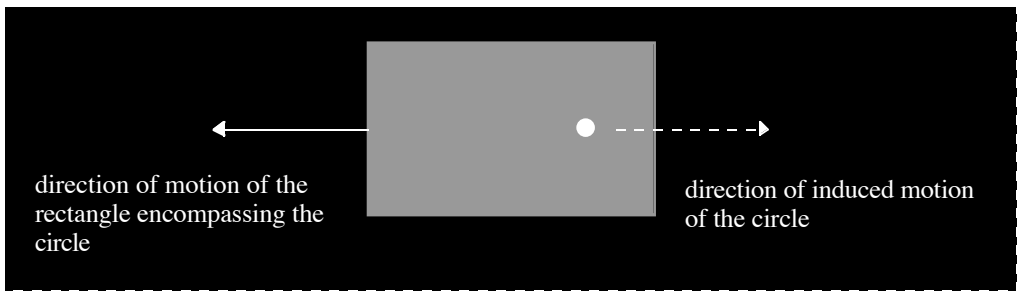

Figure 4: Induced Motion of Circle Based on Motion of the Rectangle

Where an object, such as the rectangle, that is "actually" moving is perceived to do so, the experienced velocity of the circle is about double the case when induced motion is absent (Duncker, 1929/1938). Duncker explained that this phenomenon is due to each object existing relative to its own reference frame. Thus, the motion of the circle's reference frame (i.e., the rectangle) is determined relative to its own reference frame, and the motion of the circle is determined relative to the rectangular reference frame.

---

[4] The term *actual* is in quotes because in the special theory of relativity motion is relative to a particular inertial reference frame, a reference frame in which Newton's law of inertia holds. The term *actual* connotes an absolute or preferred reference frame. As will be discussed, if an absolute or preferred inertial reference frame existed, the special theory would not hold.

[5] Rock (1975/1997) discussed the actual as opposed to only the perceived motion of an object. Rock (1975/1997), and Duncker (1929/1938) in the condensed version of his paper as well, do not note that actual movement may actually only be motion relative to the observer's reference frame, where *reference frame* is defined as the term is used in physics. Since actual motion of an object is not tied to the inertial reference frame of an observer for Rock and Duncker, there is the implication that the reference frame, in the sense used in physics, is not essential to the object's motion. This circumstance approximates the concept in physics of motion in absolute space, space that has an absolute reference frame.

[6] Rock cited a paper by Oppenheimer that discussed factors concerning whether an object served as a frame of reference. The citation: Oppenheimer, F. (1934). Optische Versuch über Ruhe and Bewegung. *Psychologische Forschung*, 20, 1-46.

*Induced Self-Motion*

Rock argued that induced motion for an object can apply to the observer himself. That is, induced self-motion can occur. Induced self-motion occurs where some part of the environment serves as a reference frame relative to the observer who is the object in which motion is induced. An example of induced self-motion may be found where an observer is sitting in his car that is itself stationary in an automatic car wash where the large washing brushes pass over the car. The observer feels himself and his car in motion as the brushes pass over the car. Rock reports another example where an individual sitting near a window on a train watches a train on a nearby track move by. The individual's visual field is taken up mostly by the passing train. In this circumstance, his own train may seem to be moving. Rock also reported that a person sitting in a car that is stopped at a traffic light and watching another car that is rolling may experience his or her car also rolling.

Duncker reported on experiments where light patterns appeared upon a screen in a dark room. In one experiment, a rectangular light pattern (36 x 39 cm.) and a circle (2 cm. in diameter) were the light shapes. The circle was enclosed by the rectangle. The observer sat a distance of 30 to 100 cm. from the screen.[7]

> Here the point or dot was always at rest, objectively, while the rectangle was given objective motion. When the subject now fixated the point he experienced the feeling of being moved with it to and fro. In some cases this was so strong as to cause dizziness. The subject had himself become a part of the induced motion system and was (phenomenally) "carried along with it". When, however, something of the room-environment was visible, the subject usually did not have this experience, for now his own phenomenal status of rest was too stable, and in this case only the rectangle was seen to move. (Duncker, 1929/1938, p. 165)

---

[7] In the orignal paper, the distance is stated "100-30 cm." (p. 165). Fractional amounts are designated in the article by the use of a ".".

# On the Support

Rock (1968) conducted an experiment where an individual seated on a stool on a wagon was moved in a straight line at a constant velocity parallel to a wall 3 ft. away in a dark room.[8] As the subject moved on the wagon, stationary small circles of light 2.5 in. in diameter 5 ft. from one another appeared on the wall about at eye-level for the subject. The circles became visible one at a time to the observer because cardboard shields extended out from the wall in front of each of the circles. The subject experienced the circles of light as moving and himself as stationary. When a pattern of luminous lines (88 in. long, 0.5 in. wide and 2 to 3 ft. from one another) appeared on the wall instead of the circles of light, the subject experienced himself as moving and the pattern of luminous lines as stationary. All of the lines in the array could be seen at the same time by the subject as he moved in the wagon, and the lines did not illuminate the room in which the experiment occurred. Different subjects were used in the two stimulus conditions. Rock expected both of the results that were obtained. With regard to the luminous lines, the columns essentially formed a reference frame that essentially dominated the observer's visual perception. The observer concluded that the reference frame was at rest and that the observer, really as an object, was moving.

> It was expected that the displacement of this array of lines, which filled much of the visual field, would suffice to generate induced movement of the self. That being the case O should veridically experience himself as moving and perceive the stimulus-array as stationary. (Rock, 1968, p. 264)

Rock reasoned that in the absence of the luminous lines as a reference frame prominent in his visual perception and in the absence of cues from his body that he was moving, the observer would not perceive himself as moving. He would perceive himself to be at rest. Rock concluded this motion of the observer induced even though the observer is in motion.

> From this [experiment] it follows that in many if not most instances of transportation in a vehicle, perception of one's self and vehicle as moving is a function of the inducing effect of the

---

[8] The observer was not given information that he was moving. Further, he was told that the bumpiness he felt and the noise he heard were due to some instruments. Instead, the bumpiness was due to the wagon moving and the noise was from the engine providing the power to move the wagon with the observer. The results of the experiment support the thesis that an observer's perception of his state of motion depends on his interpretation of sensory cues.

> displacing field that surrounds the observer. This is a paradoxical conclusion–possibly confusing to the reader–because the observer is in fact moving. Yet, as the experiment cited shows, he would not perceive this were it not for the sight of the displacing environment. (p. 236)[9]

Now it so happens that the wagon traveling at a uniform velocity in a straight line (i.e., rectilinear) resembles an inertial reference frame. An inertial reference frame is a reference frame, as defined in physics, in which the law of inertia holds. This law states that in the absence of an external force, an object is at rest or moves in a straight line. The special theory requires an observer to be at rest in an inertial reference frame (i.e., stationary). Otherwise, the special theory does not hold, as will be discussed in greater detail. Rock also theorized:

> It is not beyond the realm of plausibility that even in the case where the observer is more actively moving, as in walking, or where he is passively moving but accelerating, the inducing effect of the displacing environment plays a contributory role in the perception of the self as moving and the environment as stationary. (pp. 236-237)

It appears that Rock is arguing, in the case of passive acceleration, vestibular and proprioceptive effects need to be discounted in order for the observer to maintain that he is not moving. General relativity supports this point concerning passive acceleration. It is interesting that Rock came to this conclusion given experimental evidence that such discounting might not occur when the observer accelerates. In a third experimental scenario using a wagon to transport an observer and where there were two illuminated circles on a black background, acceleration of the observer was generally experienced by the observer as he himself accelerating and the circles that he passed as stationary (Rock, 1968).

---

[9] Rock could have argued differently, though not as convincingly. Besides citing the phenomenon of induced self-motion in the example of the passing trains, Rock cited a variant of this circumstance where the individual is sitting across the aisle from the window he is looking out of onto the passing train. Here, the individual's own train compartment fills most of the his visual field. In this case, Rock noted that the other train would exhibit induced movement. Rock though did not see this latter circumstance as fundamental, that is as reflective of the fundamental experience of the individual as regards his basic state of motion. As quoted above, Rock wrote concerning his experiment that in the absence of the displacing pattern of luminous lines, the individual does not experience induced self-motion. This means that without the external information, the individual sees himself at rest, even though he is riding in a wagon. How Rock's argument supports the special theory of relativity will be shown. Also, Rock's basic notion that the fundamental experience of the individual is that he is at rest is supported by the empirical evidence that supports the special and general theories of relativity.

Rock's scenario for acceleration, though, did not take into account a situation where the sensory cues factors due to acceleration could be discounted by another *interpretation* of the entire experimental environment. The principle of equivalence of the general theory of relativity provides this method of discounting and is the basis for the derivation of the general principle of relativity. The principle of equivalence, to be discussed, provides the theoretical basis for the discounting that is critical to the extension of Rock's thesis to what are uniformly accelerating reference frames. In the general theory of relativity, an observer's being at rest in a uniformly accelerating reference frame is central to the integrity of the general theory. Special relativity, and then general relativity, will be discussed. Before doing so, further discussion is warranted on how the relativity of motion is central to Rock's analysis.

*The Relativity of Motion in Rock's Analysis*

That an observer determines that he is or is not moving (and how he does so) is central to his perception that something is moving relative to him, as opposed to his moving relative to a stationary object (Rock, 1975/1997). One important factor that frequently underlies the change in location central to perceiving motion is displacement of the object's retinal image. Displacement of the retinal image results in the observer's perception of the object as moving if the observer, including his head and eyes, are basically stationary. Yet if the displacement of the retinal image is due to the observer turning his head for example, the object is perceived as stationary because the observer relies on cues from the movement of his head to discount displacement of the retinal image of the object.

Rock argued that in the absence of the supporting cues from one's body (i.e., from the head, eyes, or the entire body) that he is moving and where an object's retinal image is displaced, the object is seen to be moving. In Rock's analysis, it does not matter whether an observer is actually moving with uniform rectilinear velocity or is instead at rest. As regards the perception of an object's motion, he argued that induced motion has its foundation in the *relative* distance between the objects, thus allowing that a "stationary" object may be

# On the Support

seen as moving. Rock (1975/1997) argued more generally that it may be the relative displacement of one object from another, as opposed to distance between them, that is the basis for the perception of motion.

Consider an observer stationary on the earth seeing an object moving in a direction opposite to the earth's rotation around its axis and with magnitude of velocity of equal to that of the earth's. The observer actually moves very quickly (about 460 meters per second at the equator) in the earth's rotation about its axis. The object, though moving relative to the earth, is stationary relative to a reference frame anchored to the earth's axis of rotation. The perception of motion as relative and dependent on sensory cues is supported by physical analysis.

As will be shown, the relative nature of motion perception dependent on sensory cues has distinct implications for the special theory of relativity. This point allows an observer to be at rest in an inertial frame of reference where in the absence of external cues that he is moving, an observer sees himself at rest and objects that are in motion relative to him in motion. If this were not the case, it would be possible for an individual to see himself as moving in the absence of cues that he is moving. If this were the case, there could be a preferred inertial reference frame. The significance that there is no preferred inertial reference frame will be discussed shortly.

### THE SPECIAL THEORY OF RELATIVITY

A fundamental principle of the special theory is that the description of physical phenomena must be the same from different inertial reference frames in uniform translational motion (i.e., rectilinear and without rotation) relative to one another. This principle depends on observers being at rest in their respective inertial reference frames. If observers were not at rest in their respective inertial reference frames, it would be possible to determine a preferred inertial reference frame, and the special theory would not hold. What would the special theory be unable to explain? Maxwell laws of electromagnetism, for example, as well as mass-energy equivalence.

Feynman, Leighton, and Sands (1963) exemplified this fundamental principle of the special theory in terms of a light clock (i.e., a clock relying on the periodic motion of light between two points to determine its beat) at rest in one or another of two inertial reference frames in uniform translational motion relative to one another. Feynman *et al.* noted that other types of clocks besides light clocks at rest in one or the other of the inertial reference frames would also have to keep the same time as the light clocks that are at rest in the same inertial reference frame as a particular non-light clock. Why? Because if this were not the case, one could then possibly distinguish between the two inertial reference frames as regards description of the physical world.

*Feynman et al.'s Thought Experiment*

Feynman *et al.* described a hypothetical situation where a space ship moving at a uniform translational velocity is one inertial reference frame and an observer outside the space ship is in another inertial reference frame.[10] Besides light clocks, there are also human observers at rest in one or the other of the two inertial reference frames.

First, regarding light and non-light clocks that are moving with a uniform translational velocity relative to one another, Feynman *et al.* (1963) wrote:

> Not only does this particular kind of clock [i.e., the light clock] run more slowly, but if the theory of [special] relativity is correct, any other clock, operating on any principle whatsoever, would also appear to run slower, and in the same proportion– we can say this without further analysis. Why is this so?

Then Feynman *et al.* discussed light clocks and non-light clocks in their thought experiment.

> To answer the above question, suppose we had two other clocks made exactly alike with wheels and gears, or perhaps based on radioactive decay, or something else. Then we adjust these clocks so they both run in precise synchronism with our first [light] clocks [which are originally all at rest in the inertial reference frame outside the space ship]. When light goes up and back in the first [light] clocks and announces its arrival [back at the light source] with a click, the new [non-light] models also complete some sort of cycle, which they simultaneously announce by some doubly coincident flash, or bong, or other signal. One of these clocks is taken into the space ship [the "moving" inertial reference frame] along with the first kind [i.e., the light clock]. Perhaps *this* clock will not run slower, but will continue to keep the same time as its stationary counterpart, and thus disagree with the other moving clock [the light clock]. Ah no, if that should happen, the man in the ship could use this mismatch between his two clocks to determine the speed of his ship, which we have been supposing is impossible [because of the principle of relativity]. *We need not know anything about the machinery* of the new clock that might cause the effect–we simply know that whatever the reason, it will appear to run slow, just like the first one.

---

[10] Strictly speaking, without setting other conditions, neither the space ship or outside the space ship are inertial reference frames.

Now if all moving clocks run slower, if no way of measuring time gives anything but a slower rate, we shall just have to say, in a certain sense, that time itself appears to be slower in a space ship. All the phenomena there–the man's pulse rate, his thought processes, the time he takes to light a cigar, how long it takes to grow up and get old–all these things must be slowed down in the same proportion, *because he cannot tell he is moving* [italics added]. The biologists and medical men sometimes say it is not quite certain that the time it takes for a cancer to develop will be longer in a space ship, but from the viewpoint of a modern physicist it is nearly certain; otherwise one could use the rate of cancer development to determine the speed of the ship! (p. 15-6)[11]

---

[11] Essentially, Feynman et al. invoked the postulate of special relativity that the laws of physics are the same in inertial reference frames in uniform translational motion relative to one another to *account* for the lack of difference between light clocks and no-light clocks. Feynman et al. wrote:

> The principle of relativity was first stated by Newton, in one of his corollaries to the laws of motion: "The motions of bodies included in a given space are the same among themselves, whether that space is at rest or moves uniformly forward in a straight line." This means, for example that if a space ship is drifting along at a uniform speed, all experiments performed in the space ship and all the phenomena in the space ship will appear the same as if the ship were not moving, provided, of course, that one does not look outside. (p. 15-1)

> Feynman et al. noted that the change in the principle of relativity found in the special theory, as opposed to its formulation in Newtonian mechanics, is:

> *all physical laws* [including those for electromagnetism] should be of such a kind that they remain unchanged under a Lorentz transformation [instead of a Galilean transformation]. In other words, we should change, not the laws of electrodynamics, but the laws of mechanics. (p. 15-3)

# On the Support

At the beginning of this experiment, Feynman *et al.* implied that the principle of relativity states that an observer can only be at rest in his inertial reference frame. Concerning a light clock, they wrote:

> Before the man took it [a light clock] aboard, he agreed that it was a nice, standard clock, and when he goes along in the space ship he will not see anything peculiar. If he did, he would know he was moving–if anything at all changed because of the motion [of the space ship], *he could tell he was moving. But the principle of relativity says this is impossible in a uniformly moving system, so nothing has changed* [italics added] (15-6)

If periodic phenomena (that make up the timing mechanism of clocks) in an inertial reference frame other than those in the light clock mechanism did not follow the time maintained by the light clock, it is possible there would be temporal differences in these inertial reference frames such that the laws of physics would not hold in all such frames. It is central to the special theory to maintain the integrity of this postulate of relativity. There is no identifiable physical basis for this phenomenon of other periodic occurrences holding to the concept of time rooted fundamentally in the relativity of simultaneity that Einstein developed using the motion of light.

### THE GENERAL THEORY OF RELATIVITY

Though we have been exploring the special theory, the general theory of relativity also provides evidence that an observer is at rest in his reference frame, that an observer in principle does not move in a reference frame from which he measures the motion of objects. The general theory is supported by empirical data and is considered a central theory in modern physics. The general theory is based on the premise that a uniformly accelerating observer can also be seen as in an inertial reference frame experiencing a uniform gravitational field. The description of physical phenomena from the uniformly accelerating reference frame associated with the observer is equivalent to that for the inertial reference frame in a uniform gravitational field (Einstein, 1916/1952; 1922/1956; 1917/1961) and Einstein and Infeld (1938/1966). It is also possible that an observer in free fall in an inertial reference frame experiencing a uniform gravitational field can be seen to be in a local gravitation-free inertial reference frame. The following example from Einstein illustrates this last possibility.

# On the Support

Consider a small area of the earth an inertial reference frame in a uniform gravitational field. For an elevator freely falling to the earth and thus accelerating in a uniform manner, one can consider the inside of the freely falling elevator to be a local (i.e., over a small area) inertial reference frame in which the laws of physics hold. For an observer in the elevator, all of Newton's laws of motion would hold so long as objects did not come into contact with the walls of the elevator (Einstein & Infeld, 1938/1966). Thus, the supposedly absolute character of the uniformly accelerated motion of the elevator does not affect the validity of Newton's laws for the man inside the elevator. Importantly, in terms of Newton's laws of motion, an observer inside the elevator is affected by these laws no differently than an observer outside the elevator at rest on earth for the elevator and the observer in it are moving at the same velocity as they are accelerating at the same rate.

As an example of an inertial reference frame in a uniform gravitational field being equivalent with regard to the motion of bodies to a uniformly accelerating frame of reference, Einstein (1917/1961) discussed a man in a large chest resembling a room and without windows deep in outer space that is accelerating uniformly due to its being towed by a rope connected to one end of the chest. According to Einstein, the man in the chest would consider himself in an inertial reference frame in a uniform gravitational field. The man in the chest would feel the same pull toward the floor that he would in such an inertial reference frame. Also, if he dropped an object, it would fall to the floor of the chest in the same manner as if the chest were in an inertial reference frame in a uniform gravitational field. It would accelerate in a uniform manner. Essentially, Einstein asked, "What is so special about inertial reference frames when one cannot even distinguish them in certain instances from other reference frames?" His answer was, "Nothing," and thus the laws of physics should hold for all frames of reference, not just inertial ones.

Consider Einstein's own description of the experience of an observer who is at first in an inertial reference frame *and then* in an accelerating reference frame, one that in principle need not be uniformly accelerating. The example concerns a train traveling with uniform translational velocity along an embankment and then experiencing a non-uniform motion, the application of the brakes. Einstein wrote:

> It is certainly true that the observer in the railway carriage experiences a jerk forwards as a result of the application of the brake, and that he recognises in this non-uniformity of motion (retardation) of the carriage. But he is compelled by nobody to refer this jerk to a "real" acceleration (retardation) of the carriage. He might also interpret his experience thus: "*My body of reference (the carriage) remains permanently at rest. With reference to it, however, there exists (during the period of application of the brakes) a gravitational field which is directed forwards and which is variable with respect to time. Under the influence of this field, the embankment together with the earth moves non-uniformly in such a manner that their original velocity in the backwards direction is continuously reduced.*" [italics added] (Einstein, 1917/1961, pp. 69-70)[12]

Einstein had a way to develop a system to measure space and time, really space-time, in accelerating reference frames and inertial reference frames experiencing gravitational fields. What Einstein did was to break up a uniformly accelerating reference frame into a sequence of very tiny reference frames. These tiny reference frames are moving in uniform translational motion relative to one another. Then the space-time relationship between these tiny reference frames is the same as that between inertial reference frames moving in uniform translational motion relative to one another in special relativity. Einstein could develop a space-time metric for an accelerating reference frame or an inertial reference frame experiencing a gravitational field by considering the pattern of space-time relationships of these tiny reference frames in relation to one another in terms of the special theory These tiny reference frames would then be described by a new non-Euclidean metric. Some other well-known physicists commented on the importance of special relativity to general relativity. They wrote:

> General relativity is built on special relativity. (Misner, Thorne, & Wheeler, 1973, p. 164)

Elaborating on this statement, they said:

> A tourist in a powered interplanetary rocket feels "gravity." Can a physicist by local effects convince him that this "gravity" is bogus? Never, says Einstein's principle of the local [over a small area] equivalence of gravity and accelerations. But then the physicist will make no errors if he *deludes* [italics added]

---

[12] One can see the extent to which Einstein maintained an observer would discount cues in the environment (i.e., external cues) and see himself at rest.

himself into treating true gravity as a local illusion caused by acceleration. Under this delusion, he barges ahead and solves gravitational problems by using special relativity: if he is clever enough to divide every problem into a network of local questions, each solvable under such a delusion, then he can work out all influences of any gravitational field. Only three basic principles are invoked: special relativity physics, the equivalence principle, and the local nature of physics. They are simple and clear. To apply them, however, imposes a double task: (1) take space-time apart into locally flat pieces (where the principles [of the special theory] are valid), and (2) put these pieces together again into a comprehensible picture. To undertake this dissection and reconstitution, to see curved dynamic space-time inescapably take form, and to see the consequences for physics: that is general relativity. (p. 164)

Einstein proceeded to develop physical law so that it would also encompass accelerating reference frames and reference frames experiencing gravitational fields. He also knew that he could provide a physical basis for an object's gravitational mass being essentially equal to the resistance of this object to acceleration by a force. This equality is important since Galileo found that the acceleration of objects in the earth's gravitational field was the same, regardless of the type of object or its gravitational mass. That the man in the room accelerating in deep space could describe the physical world as well as someone outside watching the room accelerating requires that this equality hold. The equivalence of their descriptions forms the foundation for the general principle of relativity and the across-the-board application of the laws of physics in different types of reference frames.

It has been shown that relativity theory supports Rock's thesis that even given proprioceptive and vestibular cues associated with an observer's passive acceleration, the observer can discount them and see himself as stationary and his environment as moving. It is very interesting that Einstein relied on what he implicitly formulated as a law of psychological phenomena to develop the general theory, namely that an observer considers his fundamental state of motion to be at rest and that various external cues are used by an observer to conclude that he is in motion.

# On the Support

Pais (1982) provided a quote from an unpublished paper of Einstein's in which he discussed how he came to believe that there should be *no* preferred reference frame for the description of physical phenomena.

> Then there occurred to me the...happiest thought of my life, in the following form. The gravitational field has only a relative existence in a way similar to the electric field generated by magnetoelectric induction. *Because for an observer falling freely from the roof of a house there exists*–at least in his immediate surroundings–*no gravitational field* [his italics]. Indeed, if the observer drops some bodies then these remain relative to him in a state of rest or of uniform motion, independent of their particular chemical or physical nature (in this consideration the air resistance is, of course, ignored). The observer therefore has the right to interpret his state as 'at rest.'
>
> Because of this idea, the uncommonly peculiar experimental law that in the gravitational field all bodies fall with the same acceleration attained at once a deep physical meaning. Namely, if there were to exist just one single object that falls in the gravitational field in a way different from all others, then with its help the observer could *realize* [italics added] that he is in a gravitational field and is falling in it. If such an object does not exist, however–as experience has shown with great accuracy–then the observer lacks any objective means of perceiving himself as falling in a gravitational field. Rather *he has the right* [italics added] to consider his state as one of rest and his environment as field-free relative to gravitation.
>
> The experimentally know matter independence of the acceleration of fall is therefore a powerful argument for the fact that the relativity postulate has to be extended to coordinate systems which, relative to each other, are in non-uniform motion. (Pais, 1982, p. 178).[13,14]

One can see in the above quote that Einstein maintained that the observer's perceiving that he is in motion is cognitive in nature and based on the interpretation of sensory cues. In the absence of those cues, the observer considers himself at rest, even if he is in an accelerating reference frame. One surmises that Einstein's law concerning an observer's being at rest in the absence of cues that he is moving is very broad, even including where the observer is active. It is difficult to believe that the special or general theory of relativity depends on a psychological distinction concerning whether an observer is actively or passively moving. Both the special and general theories depend on an observer concluding that he is at rest either in the absence of cues that he is moving or in his interpretation of sensory cues.

---

[13] The text "[his italics]" is in the original quote from Pais.

[14] The unpublished paper was written in 1919 and/or 1920 and is in the Pierpont Morgan Library in New York City.

## CONCLUSION

The existence of induced self-motion is supported the special and general theories of relativity. In particular, where the observer is moving, as determined by an expenditure of energy, the observer's experience that he is in motion may be induced. Also, Rock's analysis of the relativity of motion provides the basis for the central concept of the special theory of relativity that observers are at rest in inertial reference frames in uniform rectilinear motion relative to one another.

Einstein essentially extended Rock's thesis on induced self-motion by discussing a situation where proprioceptive and vestibular cues that could indicate that the observer is accelerating are discounted. This discounting is implicit in the principle of equivalence of the general theory of relativity. By limiting the observer's visual perception to those objects accelerating with him, Einstein concluded the observer would see himself at rest, just as observers at rest on the earth see themselves at rest.

Though Einstein and other physicists recognized the importance of an observer being at rest in an inertial reference frame for the special theory of relativity, the supporting psychological structures were not discussed much by physicists. On the other hand, Rock wrote of the factors involved in the perception of motion, including one's own. Rock thus came to discuss issues of significance to relativity theory, apparently without any significant understanding of how his theory might be related to relativity theory. In this paper, connections between Rock's theory on the perception of one's own motion, as well as empirical work supporting it, and relativity theory have been explored.

# On the Support